\documentclass[12pt,letterpaper]{article}

\usepackage{ol2}
\begin{document}

\title{Collision and fusion of counterpropagating micron-sized optical beams in non-uniformly biased photorefractive crystals}

\author{A. Ciattoni} \email{alessandro.ciattoni@aquila.infn.it}
\affiliation{Consiglio Nazionale delle Ricerche, CASTI Regional Lab 67100 L'Aquila,
Italy} \affiliation{Dipartimento di Fisica, Universit$\grave{a}$ dell'Aquila, 67100 L'Aquila, Italy}

\author{A. Marini}
\affiliation{Dipartimento di Fisica, Universit$\grave{a}$ dell'Aquila, 67100 L'Aquila, Italy}

\author{C. Rizza}
\affiliation{Dipartimento di Ingegneria Elettrica e dell'Informazione, Universit$\grave{a}$ dell'Aquila, Monteluco di Roio 67100
L'Aquila, Italy}

\author{E. DelRe}
\affiliation{Dipartimento di Ingegneria Elettrica e dell'Informazione, Universit$\grave{a}$ dell'Aquila, Monteluco di Roio 67100
L'Aquila, Italy}

\date{\today}

\begin{abstract}
We theoretically investigate collision of optical beams travelling in opposite directions through a centrosymmetric
photorefractive crystal biased by a spatially non-uniform voltage. We analytically predict the fusion of counterpropagating
solitons in conditions in which the applied voltage is rapidly modulated along the propagation axis, so that self-bending is
suppressed by the "restoring symmetry" mechanism. Moreover, when the applied voltage is slowly modulated, we predict that the
modified self-bending allows conditions in which the two beams fuse together, forming a curved light-channel splice.
\end{abstract}


Photorefraction has been a principal setting for the experimental investigation of soliton collisions \cite{Review}.  One setback
in previously reported schemes is the fact that the single solitons intrinsically self-bend along their propagation
\cite{SelfBending}.  Although this has a minimal effect on spatial soliton interaction where both beams propagate in the same
direction, it strongly affects the phenomenology when the beams counterpropagate \cite{Kip2002,CountPhot1}. For example, in
soliton head-on collisions, the main physical effects emerge when there is an extended beam spatial overlap
\cite{CountPhot2,holfoc,Cohen2002}, and this overlap is strongly reduced by self-bending that curves both beams in the same
lateral direction. A possible solution to enhance spatial beam overlap is to use specific curved trajectories obtained by
launching tilted beams \cite{CountPhot1}. Both from the fundamental and applicative perspective, it is natural to formulate an
alternative strategy allowing the increased overlap without a specific tilted alignment.  For example, a self-adaptive fusion of
two counterpropagating beams could amount to an alignment-robust splicing technology for fiber connections.

Recent studies have shown how self-bending can in fact be suppressed by the "restoring symmetry" mechanism implemented through
the use of alternating bias voltage profiles along the propagation direction \cite{wiggler1,wiggler2}. In this Letter we
theoretically consider the head-on collision of two counterpropagating optical beams travelling through a centrosymmetric
photorefractive crystal in the presence of a system of electrodes delivering a voltage profile periodically modulated along the
propagation axis. When the period of the applied voltage is much smaller than the optical diffraction length, the consequent
suppression of the self-bending of each single beam allows us to investigate the interaction of two micron-sized
counterpropagating beams whose complete spatial overlap, along a \textit{straight} line, triggers their mutual longitudinally
nonlocal interaction \cite{CountPhot1,CountPhot2}. In the situation where the two beams are mutually incoherent, we show that
soliton fusion occurs by analytically proving the existence of a two-parameter family of fully overlapping counterpropagating
solitons. In conditions where the applied voltage is slowly modulated, we numerically identify the conditions for exploiting the
longitudinal wiggling beam profile \cite{wiggler2} to achieve a robust fusion of two counterpropagating beams impinging on the
opposite crystal facets. The results indicate a self-adaptive merging of two channels, an effect that has a potential photonic
application in minimizing optical power loss associated to diffraction and misalignment in optical fiber splicing.

Consider a photorefractive crystal whose facets $x=-L_x$ and $x=L_x$ have a set of electrodes that deliver the periodical
potential profiles $-V_0 \cos( \kappa_v x)$ and $V_0 \cos( \kappa_v x)$, respectively, as shown in Fig.(1). In
Ref.\cite{wiggler2}, it has been proved that, if $I(x,z)$ is the optical intensity of the light travelling along the $z$-axis
through the crystal, the photorefractively induced refractive index change is
\begin{equation} \label{deltanphot}
\delta n = \frac{\alpha}{(I + I_b)^2} \left[\psi \cos(\kappa_v z) + \chi \frac{\partial I}{\partial x} \right]^2,
\end{equation}
where $\chi = K_B T / q$, $\alpha = -(1/2) n_0^3 g \epsilon_0^2 (\epsilon_r -1)^2$ ($T$ is the crystal temperature, $q$ is the
electron charge, $\epsilon_r$ is the relative dielectric constant at the given $T$, $n_0$ is the uniform refractive index
background, $g$ is the significant quadratic electro-optic coefficient), $I_b$ is the intensity of a reference background uniform
illumination, and $\psi = V_0 I_b / [L_x \cosh(\kappa_v L_x)]$. In the TE configuration, the complex amplitude of the
monochromatic (at frequency $\omega$) optical electric field $E(x,z)$ satisfies the Helmholtz equation $(\partial_{xx} +
\partial_{zz})E + k_0^2(n_0+\delta n)^2 E =0$ where $k_0 = \omega / c$ and $\delta n$ is given by Eq.(\ref{deltanphot}). In order
to describe head-on collision of two beams counterpropagating along the $z$-axis we set $E(x,z)= \exp(i k z) A_+ (x,z) + \exp(-i
k z) A_- (x,z)$ where $k = k_0 n_0$ and $A_+$ and $A_-$ are the slowly-varying amplitudes of the forward and backward propagating
beams, respectively. Inserting this expression for $E$ into the Helmholtz equation, in the paraxial approximation and noting that
for $\kappa_v << k$ (i.e. the period of voltage modulation is much greater than the optical wavelength) light cannot be
Bragg-matched with the periodic refractive index profile, $A_\pm$ satisfy the coupled parabolic equations $[\pm i
\partial_z + (1/2k)\partial_{xx}] A_\pm = -(k/n_0) \delta n
A_\pm$. We here focus our attention on the relevant case of two mutually incoherent counterpropagating beams for which the total
optical intensity is given by $I = |A_+|^2 + |A_-|^2$. Since Eq.(\ref{deltanphot}) is not an even function of $x$ if the
intensity $I$ is even, the two beams $A_\pm$ do not propagate on a straight line:  they experience the effect of self-bending in
the same lateral direction, so that the effect of their interaction is generally limited by the smallness of the overlapping
region. In order to maximize the overlap of the two beams we take $\kappa_v \gg 2\pi / L_d$ (the situation corresponding to the
geometry illustrated in Fig.(1a)) where $L_d$ is the longitudinal scale characterizing the propagation of $A_\pm$ ($L_d$
typically coincides with the optical diffraction length). In these conditions the optical beams are not able to follow the rapid
voltage oscillation and the averaged fields do not experience self-bending, and do not wiggle (the mechanism of "restoring
symmetry" discussed in Ref.\cite{wiggler2}). In this regime (i.e. $\kappa_v \gg 2\pi / L_d$) it is possible to set $A_\pm  =
\sqrt{I_b} V_\pm + \delta A_\pm$, where $V_\pm$ are those parts of the fields having a longitudinal scale of variation $L_d$, and
$\delta A_\pm$ are longitudinally rapidly varying, on a scale $2 \pi / \kappa_v$, and they are uniformly in the condition
$|\delta A_\pm| \ll \sqrt{I_b} |V_\pm|$. This self-consistent decomposition of the fields into a slowly varying mean-field
component and a rapidly oscillating and small correction allows us to derive a set of equations for $V_\pm$ (following a
procedure very close to that reported in Ref.\cite{wiggler2}) that are
\begin{eqnarray} \label{collequa}
i \frac{\partial V_+}{\partial \zeta} + \frac{\partial^2 V_+}{\partial \xi^2} &=& \frac{ \displaystyle \frac{1}{2} + \gamma
\left[ \frac{\partial}{\partial \xi} \left( |V_+|^2+|V_-|^2 \right) \right]^2 }{\left[1+|V_+|^2+|V_-|^2\right]^2} V_+, \nonumber
\\
-i \frac{\partial V_-}{\partial \zeta} + \frac{\partial^2 V_-}{\partial \xi^2} &=& \frac{ \displaystyle \frac{1}{2} + \gamma
\left[ \frac{\partial}{\partial \xi} \left( |V_+|^2+|V_-|^2 \right) \right]^2 }{\left[1+|V_+|^2+|V_-|^2\right]^2} V_-
\end{eqnarray}
where we have also introduced dimensionless variables according to $\xi = k |\psi/I_b| \sqrt{2|\alpha|/n_0} x$, $\zeta = k
(|\alpha|/n_0)(\psi/I_b)^2 z$ and $\gamma = 2 k^2 \chi^2 |\alpha|/n_0$. Note that, as expected, the two beams $V_\pm$ are driven
by the very same nonlinear waveguide (as a result of the two beams mutual incoherence) which, if $|V_+|^2$ and $|V_-|^2$ are
transversally even functions of $\xi$, is transversally even as well so that no self-bending occurs. Equations (\ref{collequa})
admit of the solution
\begin{eqnarray} \label{solitfusion}
V_+(\xi,\zeta) &=& \cos\Phi \exp \left[i\frac{a}{2}\xi - i\left(\frac{a^2}{4}-\beta\right) \zeta \right]  v(\xi-a\zeta),
\nonumber \\
V_-(\xi,\zeta) &=& \sin\Phi \exp \left[-i\frac{a}{2}\xi + i\left(\frac{a^2}{4}-\beta\right) \zeta \right]  v(\xi-a\zeta)
\end{eqnarray}
for any values of the real parameters $\Phi$ and $a$ if the function $v(\tau)$ satisfies the equation
\begin{equation} \label{solitons}
\frac{d^2 v}{d \tau^2} =  \beta v + \frac{\frac{1}{2}+\gamma \left( \frac{d v^2}{d \tau}\right)^2}{(1+v^2)^2} v.
\end{equation}
Note that Eq.(\ref{solitons}) coincides with the equation describing solitons propagating through the medium in the presence of
the "restoring symmetry" mechanism, as discussed in Ref.\cite{wiggler2}, so that the fields in Eqs.(\ref{solitfusion}) constitute
a two-parameter family of counterpropagating solitons. It is worth stressing that the two solitons of each pair do not suffer
self-bending, are fully overlapping and therefore Eqs.(\ref{solitfusion}) describe fusion of solitons counterpropagating along a
straight line. The parameter $\Phi$ sets the mutual power content of the two solitons in such a way that $|V_+|^2+|V_-|^2 =
|v|^2$, whereas the parameter $a$ (subjected to the restriction $a \ll 1$ required by the paraxial approximation) allows the
soliton pair to be slightly tilted with respect to the $z$-axis.

In order to check the above analytical results and to extend our investigation to the off axis interaction configuration (see
Fig.(1b)), we have integrated the full time-dependent photorefractive nonlinear optical model \cite{Photorefractive}. In our
numerical approach, at each instant of time, we evaluate the electric field distribution induced by the boundary applied voltage
and the photoinduced charge solving the $(x,z)$ electro-static Poisson equation, and the corresponding optical field distribution
determined by the electro-optic response through the parabolic equation \cite{cha-sat}. We have chosen a crystal bulk (layer) of
potassium lithium tantalate niobate (KLTN) ($n_0=2.4$) of thickness $2L_x = 2 \times 50 \: \mu m$ and length $L_z = 1000 \: \mu
m$. In order to investigate fusion of coaxial counterpropagating beams in the fast modulated regime (with electrode modulation
period $2 \pi / \kappa_v = 200 \: \mu m$), we have chosen $\gamma = 0.2$ (an experimentally available situation as reported in
Ref.\cite{DifEug}) and we have launched two identical counterpropagating Gaussian beams $A_+(x,0)=\sqrt{I_b/2} f(\xi)$ at $z=0$
and $A_-(x,L)=\sqrt{I_b/2} f(\xi)$ at $z=L$ where $f(\xi)$ is a real Gaussian profile centered at $\xi=0$ ($\xi$ is the same
dimensionless spatial variable as in Eqs.(\ref{collequa})). We have performed various numerical simulations varying both Gaussian
width and amplitude together with the applied voltage thus determining their values which maximize the overlap between the
forward and backward propagating field profile at $z=0$, so to observe the formation of a stable and straight optical channel
(see Fig.(1a)). In Fig.(2a) we have plotted the FWHM, $\sigma$, of $f^2(\xi)$ as a function of $f_0 = f(0)$ (plotted as stars)
corresponding to coaxial beam fusion and, for comparison purposes, we have also reported the theoretical soliton existence curve
(solid line) derived by Eq.(\ref{solitons}) (see Ref.\cite{wiggler2}). The good qualitative agreement between the two different
situations indicates that the fusion mechanism is robust and feasible.

If the applied voltage is slowly modulated, a form of modified self bending occurs since the optical beams are able to
adiabatically follow the electrode modulation \cite{wiggler1,wiggler2}. This property can be profitably exploited to design a
configuration where two beams, impinging onto the crystal facets $z=0$ and $z=L$ along two parallel propagation directions, are
made to form a single and curved light channel within the crystal bulk. In the circumstance of the geometry depicted in Fig.(1b),
for example, the applied voltage is reversed one time along the $z-$ axis so that beams fusion is possible since the optical
beams bend toward negative and positive $x$ direction in the $z<500 \: \mu m$ and $z
> 500 \: \mu m$ crystal regions, respectively. We have
investigated off axis beam fusion in the configuration as in Fig.(1b) by means of the above discussed numerical scheme by
launching two identical but shifted Gaussian profiles (i.e. by setting $A_+(x,0)=A_0 \exp [(x-d/2)^2)/(2s^2)]$ at $z=0$ and
$A_-(x,L)=A_0 \exp[(x+d/2)^2)/(2s^2)]$ at $z=L$, where $A_0 = 2 \sqrt{I_b}$ and $s = 3 \: \mu m$) for various different applied
voltages $V$ and mutual beam displacement $d$ thus determining their values which maximize the overlap between the forward and
backward propagating field profile at $z=0$. The result of these calculations are reported in Fig.(2b) from which we note that
fusion can be attained even for counterpropagating beams whose mutual distance $d$ is much greater than their common width. This
result suggest a feasible way for obtaining self-adaptive optical fiber splicing.

\newpage
List of Figure Captions
\begin{itemize}
\item{Figure 1: Geometry of the collision between counterpropagating optical beams (reported as shaded regions around $x=0$)
through a non-uniformly biased photorefractive crystal layer (black and gray stripes are here electrodes at opposite potentials).
(1a) Fusion of two coaxial counterpropagating solitons in the fast modulated regime. (1b) Merging of two off axis
counterpropagating beams into a single optical channel due to modified self-bending in the slowly modulated regime.}

\item{Figure 2: (2a) Intensity full width at half maximum FWHM $\sigma$ as a function of the peak amplitude $f_0$ of the Gaussian
input counterpropagating beam profiles allowing beam fusion (stars) and corresponding theoretical existence curve (solid line)
associated with counterpropagating soliton fusion (evaluated from Eq.(\ref{solitons})). (2b) Voltage V as a function of the
displacement $d$ between the two beams required to form an optimal fused splice along a curved trajectory.}
\end{itemize}

\newpage

\begin{figure}  \centering
\includegraphics[width=0.45\textwidth]{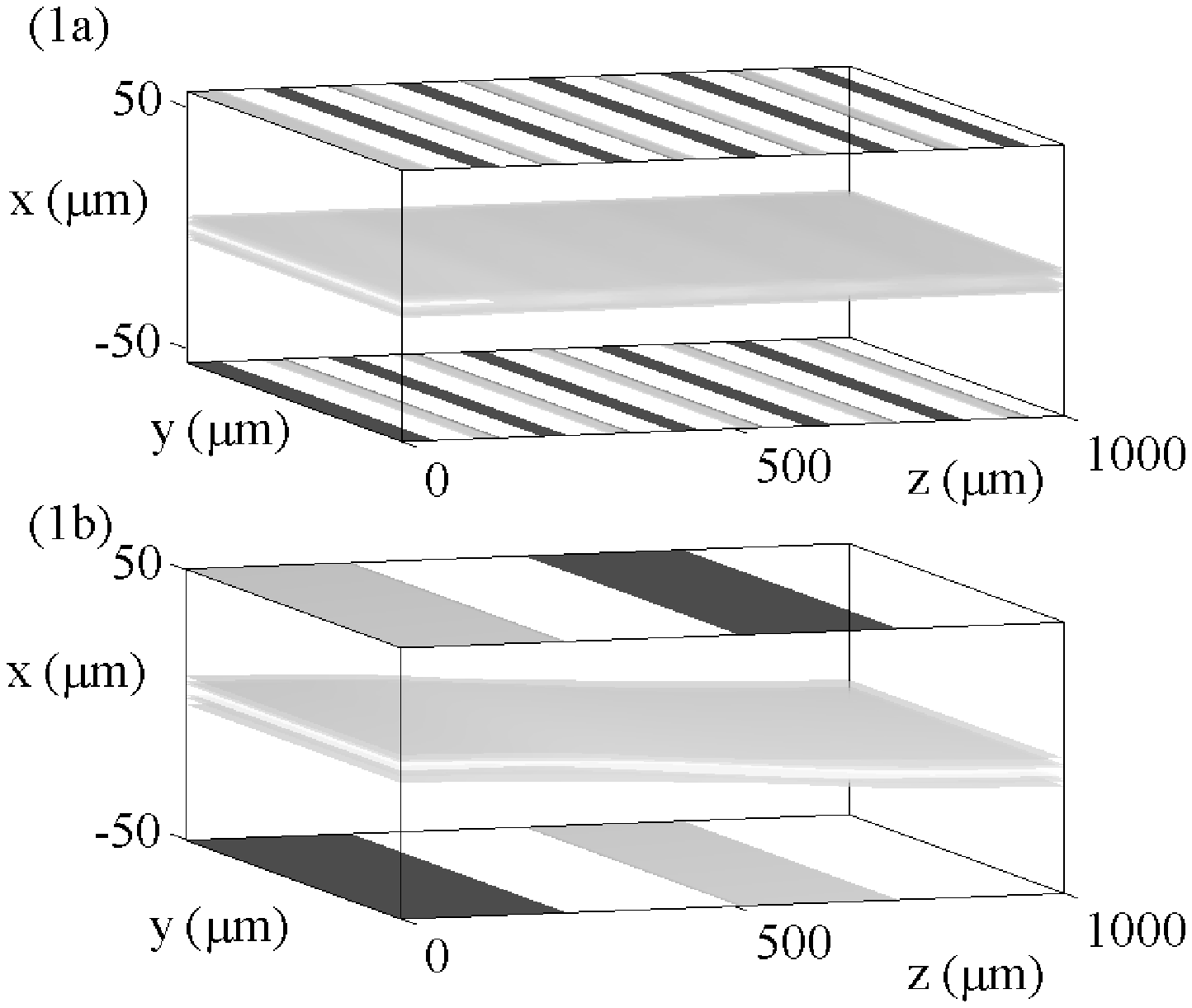}
\caption{Geometry of the collision between counterpropagating optical beams (reported as shaded regions around $x=0$) through a
non-uniformly biased photorefractive crystal layer (black and gray stripes are here electrodes at opposite potentials). (1a)
Fusion of two coaxial counterpropagating solitons in the fast modulated regime. (1b) Merging of two off axis counterpropagating
beams into a single optical channel due to modified self-bending in the slowly modulated regime.}

\includegraphics[width=0.45\textwidth]{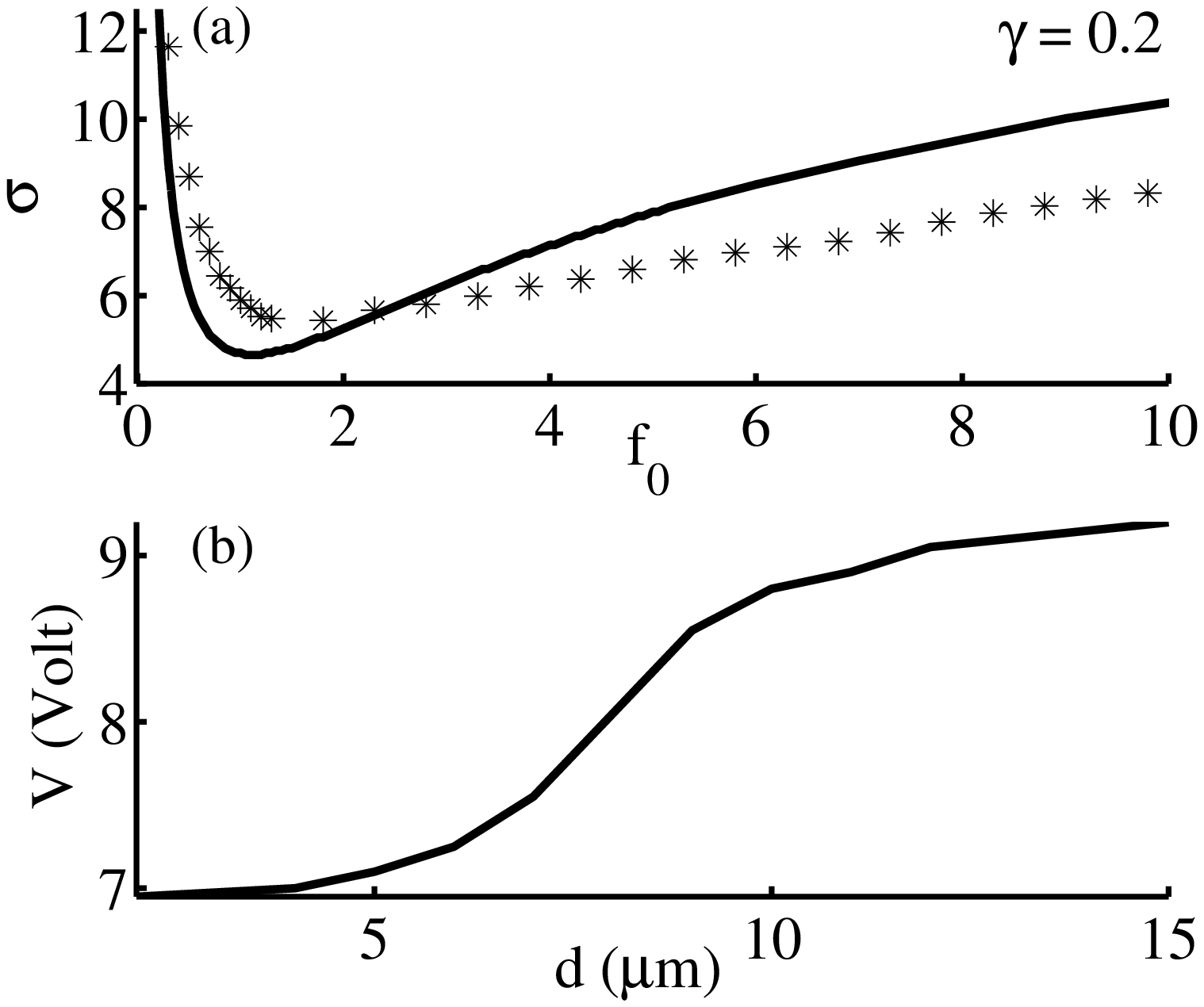}
\caption{(2a) Intensity full width at half maximum FWHM $\sigma$ as a function of the peak amplitude $f_0$ of the Gaussian input
counterpropagating beam profiles allowing beam fusion (stars) and corresponding theoretical existence curve (solid line)
associated with counterpropagating soliton fusion (evaluated from Eq.(\ref{solitons})). (2b) Voltage V as a function of the
displacement $d$ between the two beams required to form an optimal fused splice along a curved trajectory.}
\end{figure}

\end{document}